\documentclass{ws-mpla}
\usepackage{color} % to delete

\begin{document}

\markboth{Renato Higa}
{Nuclear clusters with Halo Effective Field Theory}

%%%%%%%%%%%%%%%%%%%%% Publisher's Area please ignore %%%%%%%%%%%%%%
\catchline{}{}{}{}{}
%%%%%%%%%%%%%%%%%%%%%%%%%%%%%%%%%%%%%%%%%%%%%%%%%%%%%%%%%%%%%%%%%%%

\title{Nuclear clusters with Halo Effective Field Theory}

\author{\footnotesize RENATO HIGA%\footnote{present address}
}

\address{Helmholtz-Institut f\"ur Strahlen- und Kernphysik
(Theorie), Universit\"at Bonn,\\
Nu\ss allee 14-16, 53115 Bonn, Germany\\
higa@itkp.uni-bonn.de}

\maketitle

\pub{Received (Day Month Year)}{Revised (Day Month Year)}

\begin{abstract}
After a brief discussion of effective field theory applied to nuclear 
clusters, I present the aspect of Coulomb interactions, with applications 
to low-energy alpha-alpha and nucleon-alpha scattering. 

%\keywords{Effective field theory, exotic nuclei}
\end{abstract}

\ccode{21.45.-v, 21.60.Gx}

\section{Introduction}

At low energies, few-body systems with large scattering length 
exhibit universal features (universality) that are independent of the 
interaction details. 
Some consequences are the existence of a shallow bound state and the 
Efimov effect\cite{efimov} in the two- and three-body sectors, 
respectively. These universal properties have a wide range of 
applications, from particle and nuclear to atomic and molecular 
physics\cite{eftrev1,tomio}. 

Universality has been put on a different light in the language of 
effective field theory (EFT)\cite{eftrev1,eftrev2}. 
EFT is suitable for energies with an associated Compton wavelength 
$\lambda\sim 1/M_{lo}$ that is much larger than the interaction radius 
$R\sim 1/M_{hi}$, where $M_{lo}$ and $M_{hi}$ are respectively the 
characteristic low and high momentum scales. 
The formalism allows for corrections of $O(M_{lo}/M_{hi})$, obtained 
in a systematic and model-independent way. 
For nuclear systems with $A\leq 4$, it provides a convincing explanation 
for some few-nucleon correlations, like the Phillips\cite{BHvK00,platter06} 
and Tjon\cite{PHM05} lines, as well as reliable error estimates for 
some astrophysical reactions like\cite{CSR} $n+p\to d+\gamma$. 

Technical and numerical complications arise for nuclei with $A> 4$. 
There are, however, interesting situations of halos and weakly bound 
nuclear clusters, where large simplifications can be achieved. 
The typical momentum $M_{hi}$ required to excite the core/clusters is 
much larger than the momentum $M_{lo}$ that binds the clusters altogether. 
The degrees of freedom then become the clusters 
themselves\cite{BHvK1,BHvK2,HHvK,CH}, usually 
stable nuclei like alpha particles and nucleons. 
In the following I present EFT studies for alpha-alpha ($\alpha\alpha$) 
and nucleon-alpha ($N\alpha$) interactions, which are the basic ones 
in order to build more complex systems, like the ${}^{6}$He halo nuclei, 
the first $0_+$ excited state in ${}^{12}$C (Hoyle state), or the 
$\frac{1}{2}_+$ excited ${}^{9}$Be state. 

%%%%%%%%
\section{EFT for nuclear clusters}

In halo EFT each cluster is represented by an elementary field, 
with short-range forces represented by contact interactions. 
This is a good approximation for systems whose binding (or resonance) 
energy has a compton wavelength $\lambda_B\sim 1/M_{lo}$ much larger 
than the radius $r_c\sim 1/M_{hi}$ 
of the largest cluster\footnote{
Even if these length scales are not quite well (but still) separated, 
corrections of the order $r_c/\lambda_B\sim M_{lo}/M_{hi}$ can be 
computed in a controlled and systematic way.}. 
To simplify the discussion, let us consider a system of two identical 
bosons, with mass $m_{\alpha}$, represented by a field $\phi$ and 
an $S$-wave strong interaction. The latter is described by the 
Lagrangian 
\begin{eqnarray}
{\cal L}&=&
\phi^{\dagger}\Bigg[i\partial_0+\frac{\vec\nabla^2}{2m_{\alpha}}\Bigg]\phi
-\,d^{\dagger}\Bigg[i\partial_0+\frac{\vec\nabla^2}{4m_{\alpha}}
-\Delta\Bigg]d+g\,\Big[d^{\dagger}\phi\phi+(\phi\phi)^{\dagger}d\Big]
+\cdots\,,
\label{eq:LOLag}
\end{eqnarray}
where we introduce an auxiliary (dimeron) field $d$, 
with ``residual mass'' $\Delta$, 
carrying the quantum numbers of two bosons in $S$-wave 
and coupling with their fields through the constant $g$. 
The dots stand for higher order terms in a derivative expansion. 

The form of the scattering amplitude depends on the magnitude 
of the parameters in the Lagrangian. Here we concentrate on 
the scale of $\Delta$, which determines the size of 
the scattering length $a_0$\footnote{
For simplicity, we set $g^2\sim M_{hi}^2/m_{\alpha}$ 
to get an effective range with a natural size, $r_0\sim 1/M_{hi}$.}. 
A minimal fine-tuning on $\Delta\sim M_{hi}M_{lo}/m_{\alpha}$ provides 
an unnaturally large $a_0\sim 1/M_{lo}$, corresponding to a system of 
strongly interacting bosons\footnote{A natural scale 
$\Delta\sim M_{hi}^2/m_{\alpha}$ generates $a_0\sim 1/M_{hi}$ and a 
system of weakly interacting bosons.}. 
This situation is analogous to the 
nucleon-nucleon ($NN$) case\cite{eftrev1,eftrev2}. The kinetic term 
$i\partial_0+\vec\nabla^2/4m_{\alpha}$ of the dimeron propagator is 
subleading compared to $\Delta$ and the amplitude has the form of 
the effective range expansion (ERE) formula, with the effective range $r_0$, 
shape parameter ${\cal P}_0$, and higher order terms treated in 
perturbation theory. 
However, the power counting for nuclear clusters is often more 
complicated. In $\alpha\alpha$ scattering for example\cite{HHvK}, 
an extra amount of fine-tuning ($\Delta\sim M_{lo}^2/m_{\alpha}$) 
is necessary to generate an even larger scattering length, 
$a_0\sim M_{hi}/M_{lo}^2$. 
As a consequence, the dimeron's kinetic and residual mass terms become 
of comparable order, which requires 
resummation of $r_0$ to all orders. 
This resummation is necessary to reproduce a narrow resonance at 
low energy\cite{HHvK,BHvK1,BHvK2}. 

\begin{figure}[htb]
\centerline{\psfig{file=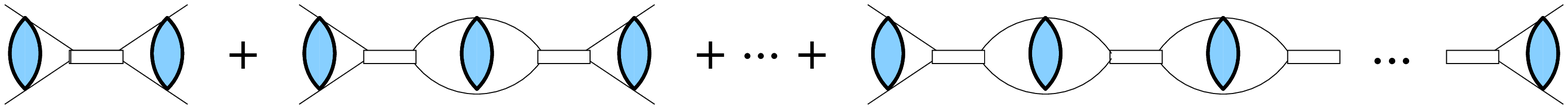,width=4.5in}}
\vspace*{8pt}
\caption{Graphic representation of $T_{CS}$, with multiple insertions of 
the bare dimeron propagator (double line) and the ``bubble loop". The 
latter contains Coulomb photons resummed to all orders (shaded ellipse).
\protect\label{fig1}}
\end{figure}
Electromagnetic interactions are introduced in the usual way, among 
which Coulomb is the dominant one at low energies\cite{HHvK}. The latter 
was formulated in the EFT framework by Kong and Ravndal\cite{clbeft} for 
the two-protons system, and can be extended in a straightforward way 
to include resonances. The Coulomb-modified strong amplitude $T_{CS}$ is 
diagramatically illustrated by Fig.~\ref{fig1} (see caption) and has 
the form of a geometric series. Like in the proton-proton case, the 
power counting for narrow resonances requires resummation. 
Up to next-to-leading order (NLO) one gets for our two-boson example, 
with reduced mass $\mu$ and charge $Z_{\alpha}$, 
\begin{equation}
T_{CS}=-\frac{2\pi}{\mu}\,C_{\eta}^2\,e^{2i\sigma_0}\,\Bigg[
\frac{1}{-\frac{1}{a_0}\!+\!k^2\,\frac{r_0}{2}\!-\!2k_C H} 
+
\frac{{\cal P}_0}{4}
\frac{k^4}{(-\frac{1}{a_0}\!+\!k^2\,\frac{r_0}{2}\!-\!2k_C H)^2} 
\Bigg]\,,
\label{eq:ampsc2}
\end{equation}
where $k_C=\mu Z_{\alpha}^2\alpha_{em}$ is the inverse of the Bohr radius, 
$\eta=k_C/k$, $C_{\eta}^2=2\pi\eta/(e^{2\pi\eta}-1)$, 
$\sigma_0=\arg\Gamma(1+i\eta)$, 
and $H(\eta)=\psi(i\eta)+(2i\eta)^{-1}-\ln(i\eta)$.
There is a small complication in this formula, that around the resonance 
energy multiple ``kinematical fine-tunings" are required\cite{BHvK2}. 
This is a technical rather than a conceptual problem, and can be 
handled via an expansion around the resonance pole\cite{HHvK}. 

%%%%%%%%
\section{Applications}

\subsection{$\alpha\alpha$ scattering}

The $\alpha\alpha$ system is dominated by $S$-wave at low energies, 
with the presence of a very narrow resonance at $E_{R}\simeq 92$~keV 
and width $\Gamma_R\approx6$~MeV (the ${}^{8}$Be ground state). 
The $\alpha\alpha$ scattering length $a_0\sim 2000$~fm is nearly 
three orders of magnitude larger than the alpha matter radius, 
suggesting the large amount of fine-tuning discussed in the last section. 
The low-energy scale $M_{lo}\sim\sqrt{m_{\alpha}E_R}\approx 
20$~MeV is roughly seven times smaller than a high momentum scale 
associated to either the pion mass or the excitation energy of the 
alpha particle, $M_{hi}\sim m_{\pi}\sim\sqrt{m_{\alpha}E^{*}_{\alpha}}
\approx 140$~MeV. 
Within the power counting for resonances, one would expect convergent 
results for observables at laboratory energies up to 3~MeV. 

Coulomb interactions are described in terms of the momentum scale $k_C$ 
which, due to the large $\alpha\alpha$ reduced mass, is numerically of 
$O(M_{hi})$. 
The low-energy amplitude is then obtained by expanding the 
function $H(\eta)$ for large $\eta$. 
However, it is interesting to discuss the other limit $k_C\to 0$, 
when Coulomb is turned off. In this case one has 
$2k_CH(\eta)\to ik\sim M_{lo}$, which is larger than the terms $1/a_0$ 
and $r_0\,k^2/2$ ($\sim M_{lo}^2/M_{hi}$). 
As a consequence, at leading order the denominator of the amplitude 
is given by the unitarity term $-ik$, and the $\alpha\alpha$ exhibits 
non-relativistic conformal invariance\cite{}. Such unitary limit 
implies the existence of the ${}^{8}$Be ground state right at 
threshold, and the corresponding three-body system (${}^{12}$C) 
showing an exact Efimov spectrum. 
These results change for a physical value of $k_C$, due to the 
breaking of conformal invariance by Coulomb forces. Nevertheless, 
the fact that the ground state of ${}^{8}$Be and the Hoyle state in 
${}^{12}$C remain very close to the threshold into $\alpha$-particles 
suggests that this conformal picture is not far from the real case. 

The low-energy expansion of $H(\eta)$ up to NLO is given by 
$1/12\eta^2+1/120\eta^4$, and resembles the usual effective range expansion. 
Defining $\tilde r_0=r_0-1/3k_C$, $\tilde{\cal P}_0={\cal P}_0+1/15k_C^3$, 
and performing an expansion around the resonance pole, $T_{CS}$ reads 
\begin{equation}
T_{CS}=-\frac{2\pi}{\mu}\,C_{\eta}^2\,e^{2i\sigma_0}\,\Bigg[
\underbrace{ 
\frac{1}{\frac{\tilde{r}_0}{2}(k^2-k_R^2)\!-\!ikC_{\eta}^2} }_{\rm LO\;\; term}
+\underbrace{ 
\frac{\tilde{\cal P}_0}{4}\,
\frac{(k^4\!-\!k_R^4)}
{(\frac{\tilde{r}_0}{2}(k^2-k_R^2)\!-\!ikC_{\eta}^2)^2}
}_{\rm NLO\;\; correction}\,\Bigg],
\label{eq:poleTCS}
\end{equation}
where $k_R$ is the momentum of the resonance. The scattering length is 
obtained from
\begin{equation}
a_0^{-1}=\tilde{r}_0\,k_R^2/2-\tilde{\cal P}_0\,k_R^4/4.
\end{equation}
An intriguing puzzle arises at this point. In the $NN$ case, 
the fine-tuning necessary to generate the low-energy bound and virtual 
states reflects a sensitivity to QCD parameters. For the present case, 
this sensitivity seems to be much larger, as indicated by two orders 
of magnitude of $\Delta$ away from a natural size. Apart from that, one 
observes an extra fine-tuning generated by a roughly 90\% cancellation 
between strong and electromagnetic contributions in the parameter 
$\tilde{r}_0$, now of $O(M_{lo}/M_{hi}^2)$. That leads to a scattering 
length of $O(M_{hi}^2/M_{lo}^3)$, an order of magnitude larger than 
its purely strong part. It is remarkable, that if the 
strong forces generated an $r_0$ 11\% larger the ${}^{8}$Be ground 
state would be bound, with drastic consequences in the formation of 
elements in the universe (see also Ref.\cite{oberh}). 

\begin{figure}[htb]
\centerline{
\psfig{file=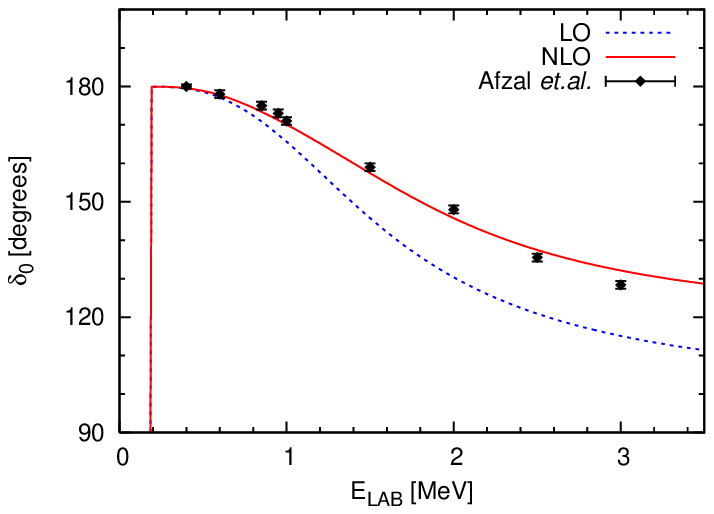,width=2.0in}
\psfig{file=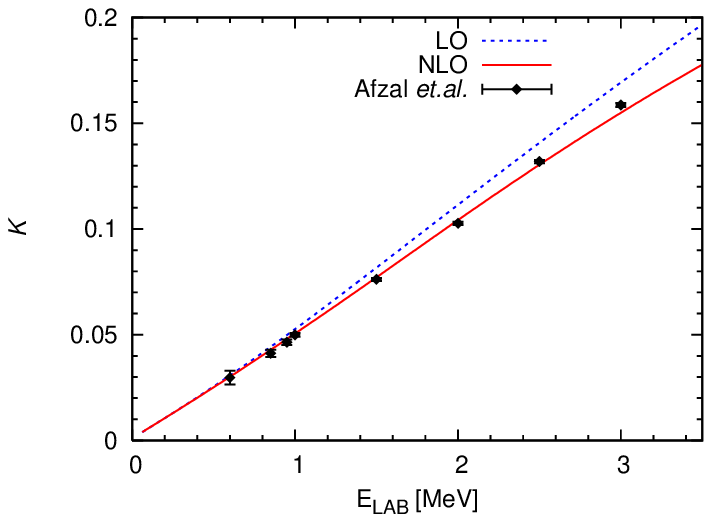,width=2.0in}
}
\vspace*{8pt}
\caption{EFT results at LO (dotted) and NLO (solid), compared against the 
scattering data. Left panel: phase shift $\delta_0^c$. Right panel: 
$K(\eta)\equiv C_{\eta}^2(\cot\delta_0^c-i)/2\eta+H(\eta)$.
\protect\label{fig2}}
\end{figure}
Despite the large fine-tunings and cancellations of strong and 
electromagnetic contributions, we obtain a successful description of 
$\alpha\alpha$ scattering at low energies, as shown in Fig.\ref{fig2}. 
$S$-wave phase shifts and ERE parameters can be found in 
a review by Afzal {\em et al.}\cite{aaa69} and references therein. 
The latest extraction of ERE parameters before our work was done in 
Ref.\cite{R67}, with numerical values given in Table~\ref{tab1}. 
We used the available scattering data, combined with the most recent 
measurement of the resonance properties\cite{Wue92}, and we were able 
to extract the ERE parameters with smaller errorbars. 
The disagreement between Ref.\cite{R67} and our results for $a_0$ 
is likely due to an approximation made in the latter, as discussed in 
details in Ref.\cite{HHvK}. 
\begin{table}[htb]
\tbl{$S$-wave effective range parameters.}
{\begin{tabular}{@{}lccc@{}} \toprule
 & $a_0$ ($10^3$ fm) & $r_0$ (fm) & ${\cal P}_0$ (fm${}^{3}$) \\
\colrule
LO & $-1.80$ & $1.083$ & --- \\
NLO & $-1.92\pm 0.09$ & $1.098\pm 0.005$ & $-1.46\pm 0.08$ \\
ERE (our fit) & $-1.92\pm 0.09$ & $1.099\pm 0.005$ & $-1.62\pm 0.08$ \\
Ref.\cite{R67} & $-1.65\pm 0.17$ & $1.084\pm 0.011$ & $-1.76\pm 0.22$\\\botrule
\end{tabular}}
\label{tab1}
\end{table}

Our combined fit, albeit showing a convergence pattern, still has a 
relative large $\chi^2/\mbox{datum}$. If one uses only the scattering 
data, the fit becomes much better but the resonance width is well 
underpredicted. This happens regardless if one uses EFT or the 
conventional ERE. We take this as indication that the measurement of 
the resonance properties and the (rather old) scattering data
are incompatible with each order or, at least, one of them has 
overestimated quoted errors. Reanalysis or even new measurements of 
scattering data seem necessary to resolve this discrepancy. 

\subsection{$N\alpha$ scattering}

At low energies $N\alpha$ scattering is dominated by the waves $S_{1/2}$ 
(or the notation $0+$), $P_{3/2}$ ($1+$), and $P_{1/2}$ ($1-$). 
A narrow, low-energy resonance is seen in the $1+$ channel, and a very 
broad one, at the $1-$ channel. 
The latter is a perturbative effect that appears only 
beyond the NLO that we are working on\cite{BHvK1,BRvK}. 
Therefore, for EFT up to this order, only $0+$ and $1+$ are relevant. 

The power counting for a narrow $P$-wave resonance was developed in 
Refs.\cite{BHvK1,BHvK2} with an application to neutron-alpha ($n\alpha$) 
scattering. Recently we incorporated the expansion around the resonance 
pole to this process, obtaining similar to (but showing better convergence 
than) the previous works\cite{BRvK}. 

\begin{figure}[htb]
\centerline{
\psfig{file=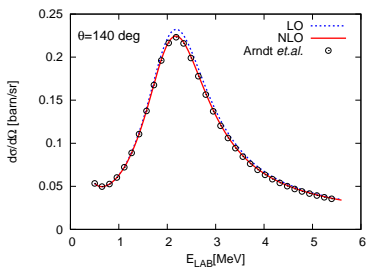,width=2.0in}
}
\vspace*{8pt}
\caption{EFT results at LO (dotted) and NLO (solid), compared against the 
partial wave analysis results from Arndt {\em et al.} (circles). 
\protect\label{fig3}}
\end{figure}
In proton-alpha ($p\alpha$) scattering, Coulomb interactions are 
included by extending the formalism from $S$-wave\cite{HHvK} to 
$P$-wave strong interactions. One important difference is that in the 
latter not only the ``scattering length'' $a_{1+}$, but also the 
``effective range'' $r_{1+}$, are renormalized by Coulomb loops. 
Fig.\ref{fig3} shows our preliminary results for $p\alpha$ 
cross-section at $\theta=140^{\circ}$ laboratory angle, compared against 
results of Ref.\cite{padat1} using the ERE. Clearly, a good agreement is 
achieved already at LO. 

%%%%%%%%
\section{Outlook}

I presented the EFT formalism for cluster resonances in the presence 
of Coulomb interactions. As applications, low-energy $\alpha\alpha$ 
and $N\alpha$ scattering were successfully described. These two interactions 
are the basic ones before considering more complicated clusters of $\alpha$ 
and nucleons. The Hoyle state in ${}^{12}$C is particularly interesting due 
to its relevance in the formation of heavy elements. A model-study with 
this state in mind was developed in Ref.\cite{limcyc}, where a perturbative 
treatment of the Coulomb interaction was proposed. This idea could be 
useful to handle the technical difficulties involving three charged 
particles. 

%%%%%%%%
\section*{Acknowledgments}

I would like to thank Hans-Werner Hammer, Bira van Kolck, and Carlos
Bertulani for stimulating collaboration, and the organizers of the 
Asia-Pacific Few-Body 2008 for the invitation and enjoyable conference. 
This work was partially support
by the BMBF under contract number 06BN411.
%, and by DOE Contract
%No.DE-AC05-06OR23177, under which SURA operates the Thomas Jefferson
%National Accelerator Facility.

%%%%%%%%

\end{document}